\documentclass[twocolumn]{aastex631}

\usepackage{threeparttable}
\usepackage{parskip}
\usepackage{float}
\usepackage{hyperref}
\usepackage{graphicx}

\usepackage{amsmath}

\usepackage{soul}

\begin{document}
\title{Long thermonuclear Burst-driven Thermal-viscous instability of accretion disk: triggering an outburst-like X-ray flare}

\author{Wenhui Yu}
\affiliation{National Astronomical Observatories, Chinese Academy of Sciences, Beijing 100101, P.R. China }
\affiliation{School of Astronomy and Space Sciences, University of Chinese Academy of Sciences, 19A Yuquan Road, Beijing 100049, China
}

\correspondingauthor{Zhaosheng Li}
\email{lizhaosheng@qut.edu.cn}

\author[0000-0003-2310-8105]{Zhaosheng Li}
\affiliation{School of Science, Qingdao University of Technology, Qingdao 266525, P.R. China}

\author{Yuanyue Pan}
\affiliation{School of Science, Qingdao University of Technology, Qingdao 266525, P.R. China}

\author{Yanan Wang}
\affiliation{National Astronomical Observatories, Chinese Academy of Sciences, Beijing 100101, P.R. China }
\affiliation{School of Astronomy and Space Sciences, University of Chinese Academy of Sciences, 19A Yuquan Road, Beijing 100049, China
}

\author{Erlin Qiao}
\affiliation{National Astronomical Observatories, Chinese Academy of Sciences, Beijing 100101, P.R. China }
\affiliation{School of Astronomy and Space Sciences, University of Chinese Academy of Sciences, 19A Yuquan Road, Beijing 100049, China
}
\begin{abstract}
We report on NICER and MAXI observations of a long-duration thermonuclear X-ray burst and a subsequent outburst-like X-ray flare from the neutron star low-mass X-ray binary MAXI J0911--655. Prior to the burst, the source was in a persistent low/hard state with a power-law-dominated spectrum ($\Gamma \sim 1.7$) and a mass accretion rate of $\sim 1\%$ of the Eddington limit. The long burst, detected by MAXI on 2020 May 22 (MJD 58991.7101), was rapidly followed up by NICER. From time-resolved spectroscopy of the cooling tail, we estimate an exponential decay time of $\approx43$ minutes, the ignition column depth of $\approx0.1\times 10^{12}~{\rm g ~cm^{-2}}$, the burst fluence of $\approx 1.1\times 10^{-4}~{\rm erg~cm^{-2}}$, and the total energy release of $\approx1.2\times10^{42}$ erg.  Approximately 1 day after the burst onset, the 0.5--10 keV light curve unexpectedly rebrightened, initiating an outburst-like flare. During the peak of this flare, the persistent power-law flux increased from its ppreburst level of $\sim0.27\times10^{-9}~{\rm erg~cm^{-2}~s^{-1}} - 1.4\times10^{-9}~{\rm erg~cm^{-2}~s^{-1}}$. This flux enhancement was accompanied by significant spectral softening, with the photon index increasing to $\Gamma \sim 2.2$. Subsequently, the flux decayed and the source returned to its baseline low/hard state. The observed timescales and energetics suggest that intense irradiation from the long burst amplified the ongoing thermal-viscous accretion process. This heating drove an inside-out heating front that temporarily enhanced the mass accretion rate, providing compelling observational evidence of a thermonuclear burst directly modulating the accretion dynamics of its surrounding disk.
\end{abstract}

\keywords{ Neutron stars (1108); X-ray bursters (1813); Low-mass X-ray binary stars (939); X-ray bursts (1814)}

\section{Introduction} \label{sec:intro}

In a low-mass X-ray binary (LMXB), a companion star ($M<1M_\odot$) transfers matter to a neutron star (NS) or black hole via disk accretion. While a small fraction of these systems are persistent X-ray sources, the majority are transients that exhibit dramatic X-ray outbursts separated by long periods of quiescence \citep{Psaltis2006}. The outbursts of transient LMXBs typically exhibit a fast-rise exponential decay (FRED) profile over typical timescales of weeks to months \citep{Lipunova_2015}. The outburst behavior can be explained by the disk instability model (DIM), in which a thermal-viscous instability, driven by the partial ionization of hydrogen (or helium, in the case of ultracompact binaries), causes the disk to cycle between a cold, neutral quiescent state and a hot, ionized outburst state \citep{Lasota01, HAMEURY20201004}.

A key ingredient in the DIM for X-ray binaries is irradiation of the disk by the central source. X-ray irradiation heats the disk, altering the stability criteria and significantly influencing the outburst properties \citep{Dubus1999, Dubus2001A&A}. Persistent irradiation can stabilize the disk against the DIM, allowing for stable accretion in persistent LMXBs, and it modifies the outburst cycles in transients by lowering the critical surface density required for an outburst to begin \citep{Lasota2008, Coban_2024}.

While persistent irradiation is a cornerstone of the DIM, LMXBs hosting NSs are subject to a more dramatic form of irradiation: powerful, brief thermonuclear (type I) X-ray bursts, resulting from unstable nuclear burning on the NS surface \citep[see][for reviews]{Strohmayer06, Galloway21}. Observations and simulations have established that the intense radiation from these bursts significantly impacts the surrounding accretion environment \citep{int2011A&A, Degenaar18, int2019A&A, Fragile20}. This interaction manifests as the ionization of the accretion disk, which produces reflection features \citep{Zhao22, Lu2023A&A, yuwh2024A&A, yuwh2025A&A}, and the cooling of the hot corona by the burst's soft photons \citep{Maccarone03, Chen18, Speicher20, Fu_2025}. Furthermore, recent X-ray and radio observations by \citet{Russell2024Natur} demonstrated that burst radiation can enhance the mass accretion rate via Poynting-Robertson (PR) drag, resulting in a brightening of the radio jet. This provides independent empirical evidence that thermonuclear bursts can directly affect the accretion flow.
Moreover, it has been suggested that this irradiation could be powerful enough to trigger a full thermal-viscous instability and a subsequent accretion outburst \citep{Kuulkers2009A&A, Serino2012}. Numerical simulations proposed that burst irradiation can raise the disk temperature by 0.5 orders of magnitude, sufficient to initiate an inside-out heating front \citep{Fragile20, Speicher2024ApJ}. However, direct observational evidence of a burst successfully triggering an outburst has remained scarce.

Most of X-ray bursts have a duration of $\sim$ 10--100 s with a typical energy release of $\sim 10^{39}$ erg, and the thermonuclear runaway starts at a typical column depth of  $y_{\rm ign}\sim 10^8 \mathrm{~g~cm^{-2}} $ \citep{Lewin93,galloway2008thermonuclear,Li18,Lu2023A&A}. A few cases of burst are longer than typical, which include intermediate-duration bursts \citep{Falanga08,Lu24}, with durations of $\sim$ 100--1000 s, and superbursts \citep{Cumming01,Li21,Peng_2025}, with the longest duration of $>10^3$ s. Most intermediate-duration bursts are powered by unstable burning of helium in a deep layer at an ignition column depth of $y_{\rm ign}\sim 10^{10} \mathrm{~g~cm^{-2}}$ \citep{Cumming06,Falanga08,Keek10}, which occur at a low mass accretion rate of $\sim 0.01\dot{m}_{\rm Edd}$ \citep{Falanga08,Alizai23}. 
Superbursts are powered by carbon burning at an ignition depth  of $\sim10^{11-12} \mathrm{~g~ cm^{-2}}$ and releasing energies of $\sim 10^{42} \mathrm{~erg}$ ~\citep{Cumming01,Strohmayer02}. Superbursts usually occurred with accretion rates higher than 10\% of the Eddington luminosity \citep{Keek12,Zand17,Li21,Peng_2025}. Observationally, however, in rarer cases of long bursts with a duration of hours were classified as intermediate-duration bursts because mass accretion rates of  $\sim1$\% of the Eddington limit cannot produce sufficient carbon to trigger a superburst \citep[see e.g.,][]{Kuulkers10,zand19A&A,Alizai23,Lu24}.

The X-ray transient MAXI J0911--655 provides an ideal laboratory to test this phenomenon. Discovered in 2016 \citep{0911_2016ATel}, it is an accreting millisecond X-ray pulsar (AMXP) in an ultracompact X-ray binary (UCXB) with a 44.3 minute orbital period, located in the globular cluster NGC 2808 at a distance of $9.45\pm{0.15}$ kpc \citep{Sanna2017A&A,Watkins2015ApJ}.  The donor star is probably a helium white dwarf with a mass of $>0.024M_{\odot}$. Since the outburst began on 2026 February 19, it has lasted about 7.8 yr.  Its long-term activity has been characterized by a persistent low luminosity ($L\approx3\times10^{36}~\mathrm{erg~s^{-1}}$; \citealt{0911_2023ATel}), corresponding to $\sim1\%$ of the Eddington limit.

In this paper, we analyze the high-cadence NICER and MAXI observations that capture a long-duration burst from MAXI J0911--655, followed by a well-resolved outburst-like X-ray flare. In Sect. \ref{sec:ob}, we introduce the observations and describe the long X-ray burst and succeeding outburst-like X-ray flare properties. In Sect. \ref{sec:Ana}, we perform the analysis of the time-resolved spectral by using NICER observations. We discuss and summarize the results in Sects. \ref{sec:Dis} and \ref{sec:sum}, respectively.

\section{Observation and Data Reduction} \label{sec:ob}

From the MAXI novae webpage\footnote{\url{http://maxi.riken.jp/alert/novae/}}, we noticed three long X-ray bursts from MAXI J0911--655, which were triggered on 2020 May 22  \citep[MJD 58991.7101;][]{2020ATel_maxi}, 2022 July 22 (MJD 59782.1030), and 2023 September 1 (MJD 60188.5488), respectively. For the first burst, NICER carried out a rapid follow-up observation (Obs. ID 3030080101) at MJD 58991.7907, 1.95 hr after the MAXI trigger, capturing the burst's cooling tail \citep{2020ATel}.

\begin{figure}[ht!]
\includegraphics[width=\linewidth]{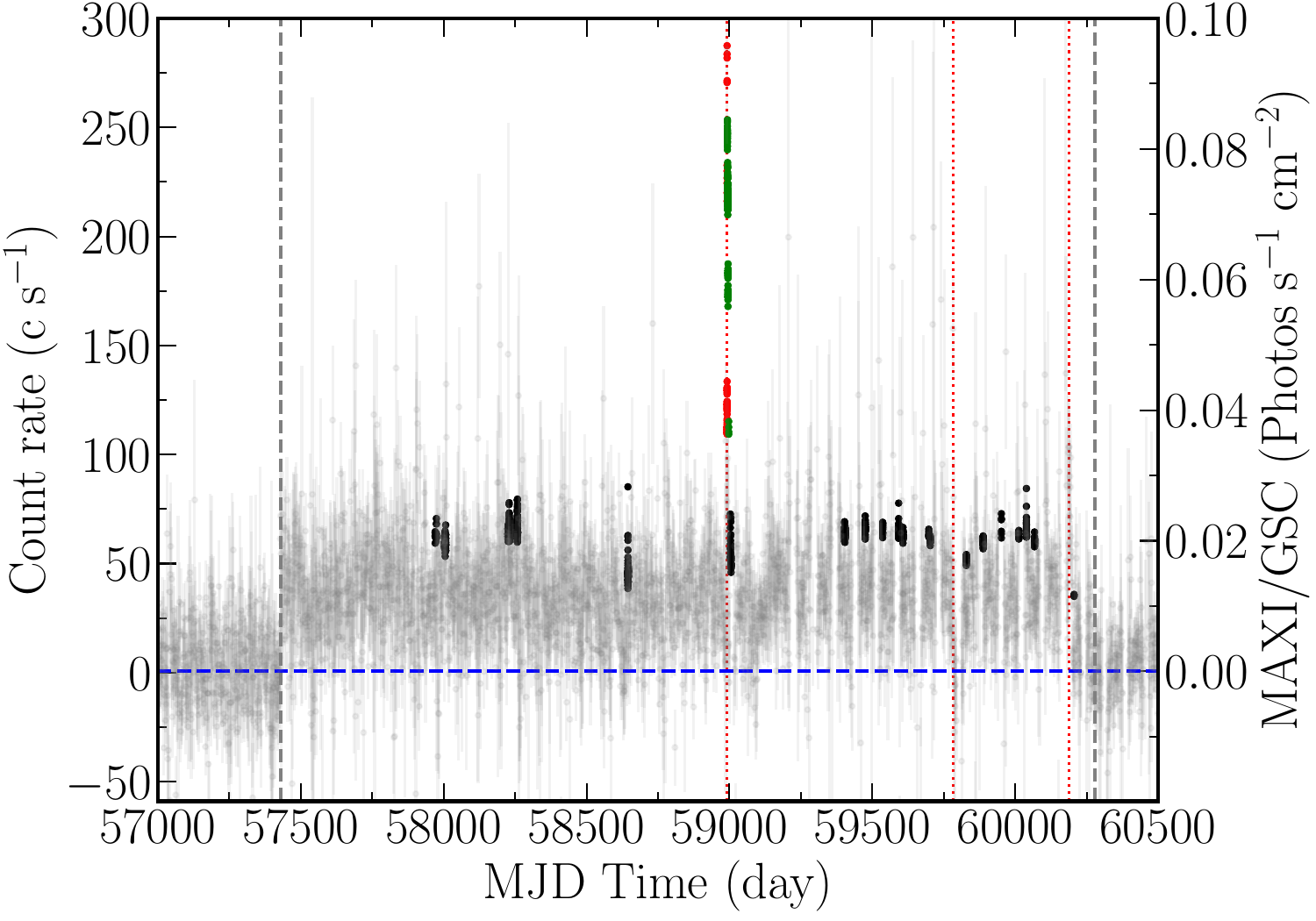}
\caption{Light curves of MAXI J0911--655 from MAXI and NICER observations. We show the  NICER light curve (64 s, 0.5--10~keV; black, red and green points) and the MAXI light curve (one day averaged, 2--20~keV; gray points), where the red and green points represent the NICER data of long X-ray burst and outburst-like flare, respectively. The vertical red dashed lines on MJD 58991.7101, 59782.1030 and 60188.5488 represent the MAXI burst trigger time. The vertical gray dashed lines on MJD 57437 and 60280 represent the beginning and end time of outburst. The horizontal blue dashed line denotes the MAXI instrumental detection threshold below which the source cannot be detected} \label{fig:lc}
\end{figure}

We analyzed all archived NICER data for MAXI J0911--655. We found 161 observations with a net unfiltered exposure time of 135.4 ks, including 11 observations (Obs. IDs 3030080101--3030080111) after the first long burst, for a total exposure time of 15.5 ks. We processed the NICER data by using HEASOFT V6.33.2 and the NICER Data Analysis Software (NICERDAS) V2.0.7. We used \texttt{nicerl3-lc} to extract the light curves in the 0.5--10, 0.5--3, and 3--10 keV bands. To construct color-color and hardness-intensity diagrams (CCDs and HIDs), we also define the soft color as the count rate ratio of 1.1--2.0~keV/0.5--1.1~keV and hard color as 3.8--6.8~keV/2.0--3.8~keV. The light curves in the energy range 0.5--10 keV and bin size of 64 s of MAXI J0911--655 observed by NICER from 2017 August to 2023 December are shown in Fig.~\ref{fig:lc}. To have a better coverage of the outburst, we added the 2--20 keV with 1 day binned light curves from the MAXI webpage \footnote{\url{http://maxi.riken.jp/mxondem/}} (see Fig.~\ref{fig:lc}).

\subsection{Light curve of long burst \label{subsec:burst_lc}}
The hardness ratio between 3--10 and 0.5--3 keV is displayed in the bottom panel in Fig.~\ref{fig:lc_burst}. The burst light curve of NICER began 1.95 hr after the MAXI trigger, marked by red points. The count rate from Obs. ID 3030080101 decreased from 230 to 100 cts s$^{-1}$ over 2.0 hr.  The hardness ratio decreased from 0.35 to 0.15, indicating the spectral soften during the burst cooling. The burst tail light curve in the energy range 0.5--10~keV can be modeled by an exponential function:
\begin{equation}
C(t)=C(t_0)e^{-(t-t_0)/\tau_{\rm LC}}+C_0. 
\label{eq:exp}
\end{equation} 
The model includes the normalization, $C(t_0)$, the exponential decay time, $\tau_{\rm LC}$, and a constant, $C_0$, representing the burst's persistent count rate. We fixed the $t_0$ to the MAXI trigger time and obtained an exponential decay time $\tau_{\rm LC}=0.72\pm0.04$ hr,($\sim 43$ minutes) suggesting that it is a long-duration burst.

\subsection{Light Curve of Outburst-like X-Ray Flare \label{subsec:outburst_lc}}
\begin{figure}[ht!]
\includegraphics[width=\linewidth]{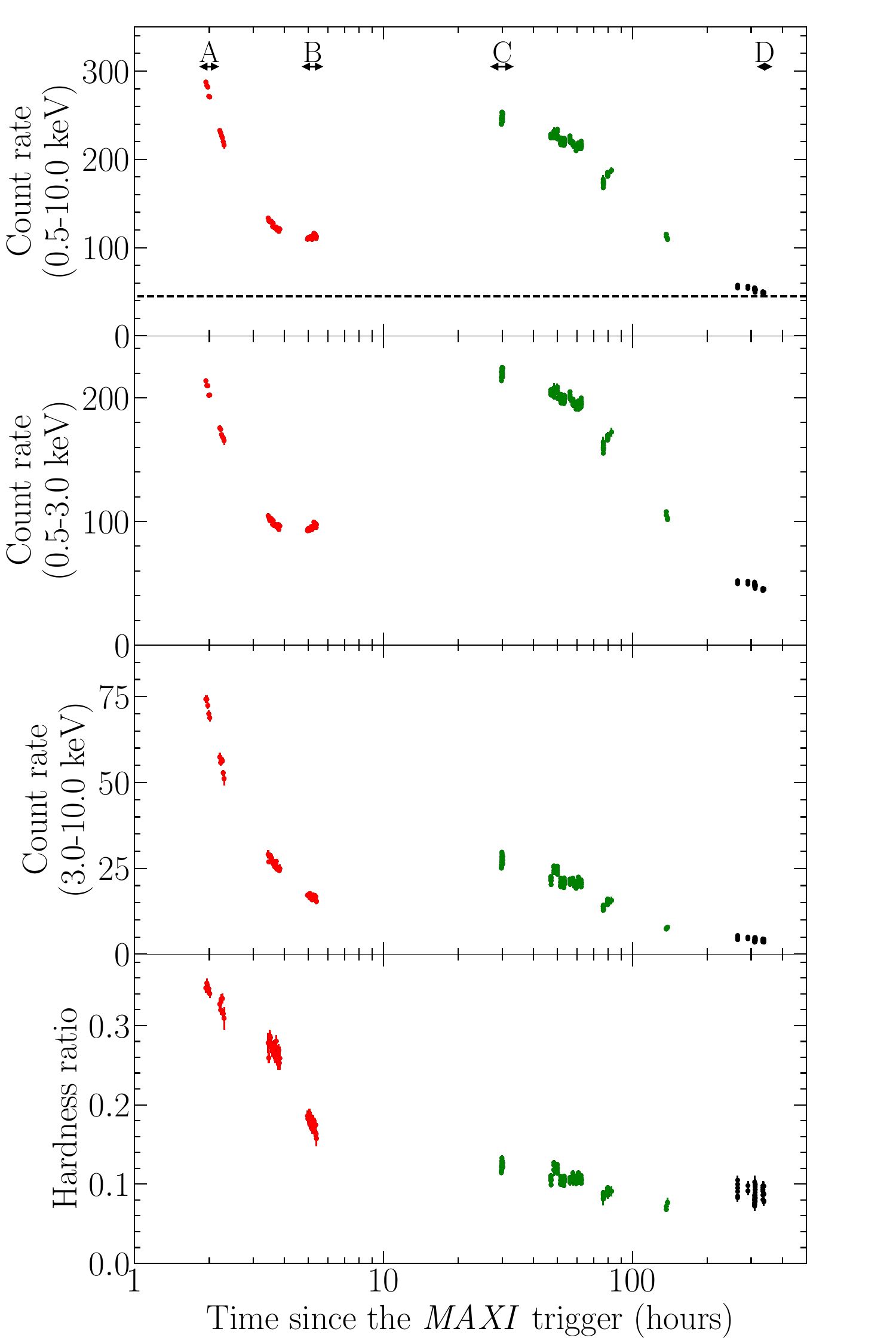}
\caption{The light curves and hardness of MAXI J0911--655 of the NICER observation between MJD 58991 and 59006. From top to bottom, we show the 64 s NICER light curves in the energy ranges 0.5--10, 0.5--3, and 3--10~keV and the hardness ratio between 3.0--10 and 0.5--3 keV, respectively.
The time intervals of the regions A, B, C, and D are represented. 
}\label{fig:lc_burst}
\end{figure}

The source had rebrightened to 250 cts s$^{-1}$ in a subsequent observation 1 day after the burst onset (see Fig.~\ref{fig:lc_burst}), well above the historic count rate of $\sim$ 50--70 cts s$^{-1}$ by  NICER \citep{0911_2019ATel_nicer,0911_2021ATel_nicer}. Subsequently, the count rate decreased from 250 to 50 cts s$^{-1}$ in 10 days.  The hardness ratio  was almost unchanged at a level of $\sim 0.1$. Based on the FRED light curve profile, we propose the rebrightened as an outburst-like X-ray flare \citep{Wood_2001,Ertan02,Lipunova_2015}.

\begin{figure}[ht!]
\includegraphics[width=\linewidth]{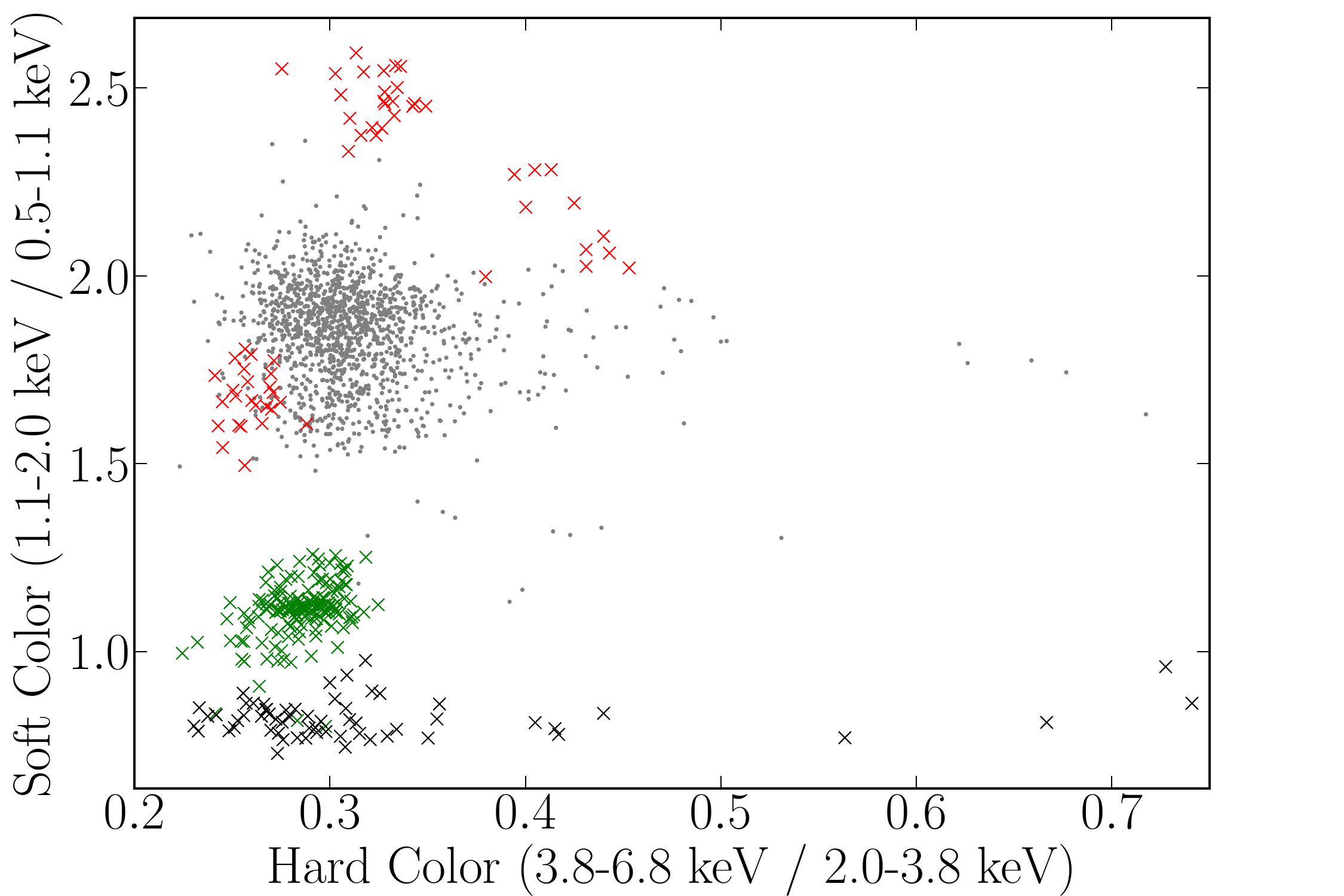}
\includegraphics[width=\linewidth]{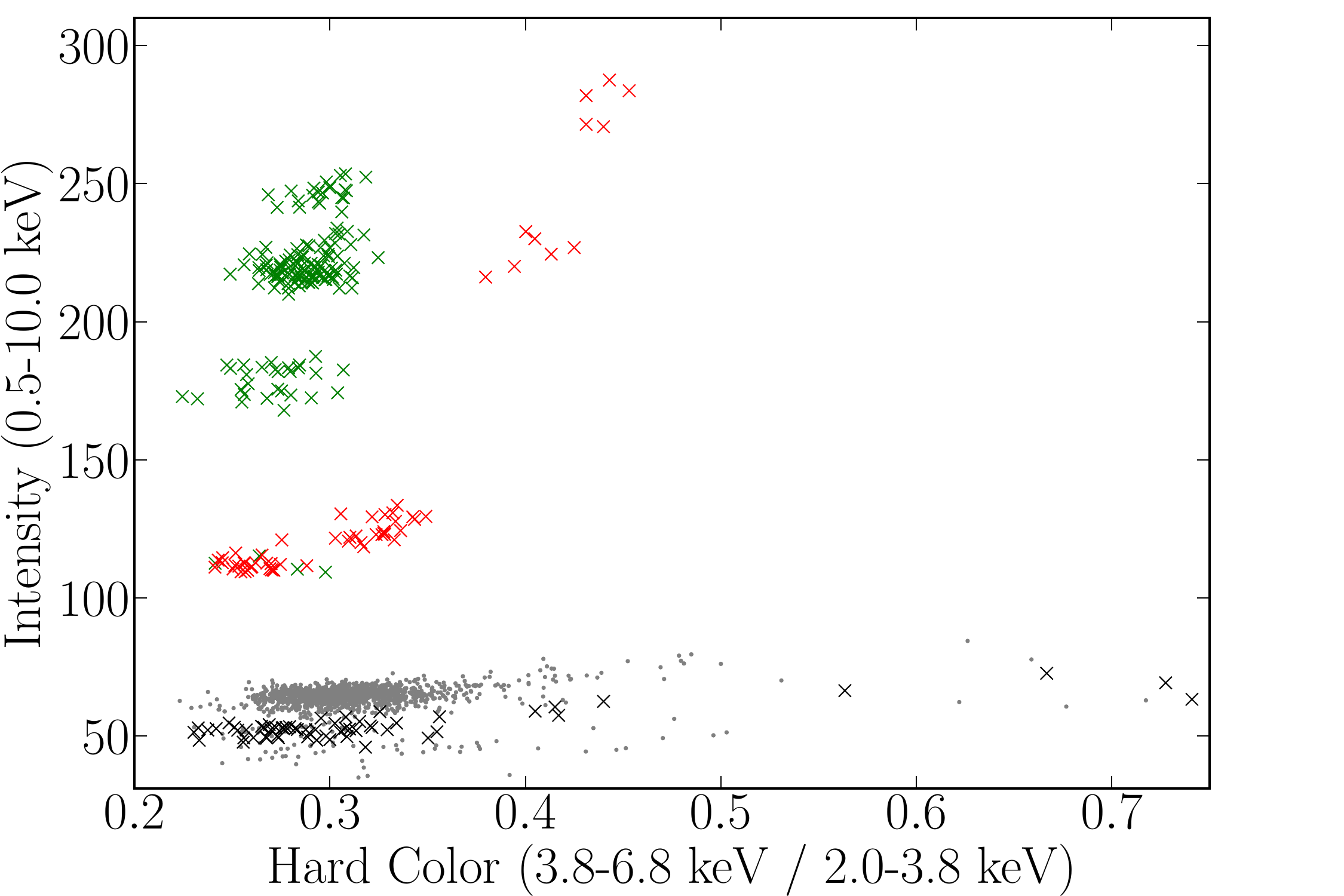}
\caption{  CCD (top) and HID (bottom) of MAXI J0911--655 from the NICER observations. The red, green, and black crosses represent the data observed in the burst, outburst-like flare, and outburst-like flare tail, respectively. Each point represents a segment of 64 s. The HID and CCD of the normal outburst observed in 2019 and between 2021 and 2023 and marked are shown as gray.}
\label{fig:HID}
\end{figure}

We show the HID and CCD of the observations between MJD 58991--59006 from MAXI J0911--655 in Fig.~\ref{fig:HID}. The soft color in Fig.~\ref{fig:HID} decreased from 1.3 to 0.8 during the flare. We noted it was lower than the normal outburst soft color $\sim~1.5-2.0$, which was observed in 2019 and between 2021 and 2023 and marked as gray points in Fig.~\ref{fig:HID}.   On the other hand, the hard color was nearly constant in the range of 0.2--0.4 in most of the time.

\subsection{Burst oscillation search \label{subsec:timing}}

As this event represents the first thermonuclear burst from MAXI J0911--655 observed with high-time-resolution X-ray data, it provides the first opportunity to search for burst oscillations in this source. Accordingly, we conducted a targeted search around its known spin frequency of $\sim$340 Hz \citep{Sanna2017A&A}. We used the barycenter-corrected event files in the 0.2--10, 0.3--3, 3--6, and 6--12 keV energy ranges. We adopted the moving window method with the width of $\Delta T=4$ s and steps of 0.5 s. We searched for a coherent signal frequency between 335 and 345 Hz  using the $Z_1^2$-test statistic \citep[also known as the Rayleigh test;][]{Buccheri83, Huppenkothen19}, which is highly sensitive to sinusoidal signals at a known frequency. The search was implemented using the  {\tt Stingray} library \citep{Bilous19, Li21}.  No statistically significant burst oscillations were detected in any individual time window during the burst's cooling tail. We placed a 99.7\% confidence upper limit on the fractional rms amplitude of any oscillation signal at 4.3\% at 340 Hz.

\section{Spectral Analysis} 
\label{sec:Ana}

For the long burst and its subsequent outburst-like X-ray flare, we extracted the time-resolved spectra with time bins of 100 s from Obs. ID 3030080101--3030080111. We extracted the spectra, ancillary response files ($\texttt{ARFs}$), response matrix files ($\texttt{RMFs}$), and the 3c50 background spectra  \citep[][]{Remillard21} using \texttt{nicerl3-spect}. We performed the optimal binning to the spectra using \texttt{ftgrouppha} as recommended by the NICER team. We performed the spectra analysis using Xspec v12.14.1 \citep{Arnaud96}. All bolometric fluxes were estimated in the energy range of 0.01--250 keV using the \texttt{cflux} model. The errors of all parameters are quoted at the $1\sigma$ confidence level.

\subsection{Spectral Evolution of the Burst Tail \label{subsec:burst}}

We performed the time-resolved burst spectral analysis for MAXI and NICER observations separately. Because the NICER observation (Obs. ID 3030080101) began during the burst's decay, a separate, contemporaneous ppreburst observation to characterize the persistent emission was unavailable. Therefore, in our time-resolved burst spectral analysis, we only subtracted the 3C50 background. The resulting spectra are thus a combination of the burst emission and the underlying persistent emission. For the MAXI data, we downloaded four time-resolved burst spectra and the corresponding background spectra and RMFs with an exposure time of 167 s from MAXI/GSC.\footnote{\href{http://maxi.riken.jp/mxondem/}{http://maxi.riken.jp/mxondem/}}
We used the model \texttt{TBabs*(bbodyrad + powerlaw)} to fit the burst spectra from NICER and MAXI, where \texttt{TBabs} accounts for interstellar absorption, with abundances set to those of \citet{Wilms_2000}. The \texttt{bbodyrad} component represents the burst emission, characterized by its temperature, $kT_{\rm bb}$, and normalization, $K = (R_{\rm bb}/D_{10})^2$, where $R_{\rm bb}$ is the apparent emitting radius in km and $D_{10}$ is the source distance in units of 10~kpc. The model \texttt{powerlaw} component accounts for the persistent emission, defined by its photon index, $\Gamma$, and normalization \citep[see e.g.,][]{Sanna2017A&A,2020ATel}{}. However, for the MAXI burst spectra, we found that the contribution of \texttt{powerlaw} can be neglected. Therefore, we only obtained the best-fitting parameters of the blackbody component from MAXI spectra.

\begin{figure}[ht!]
\includegraphics[width=\hsize]{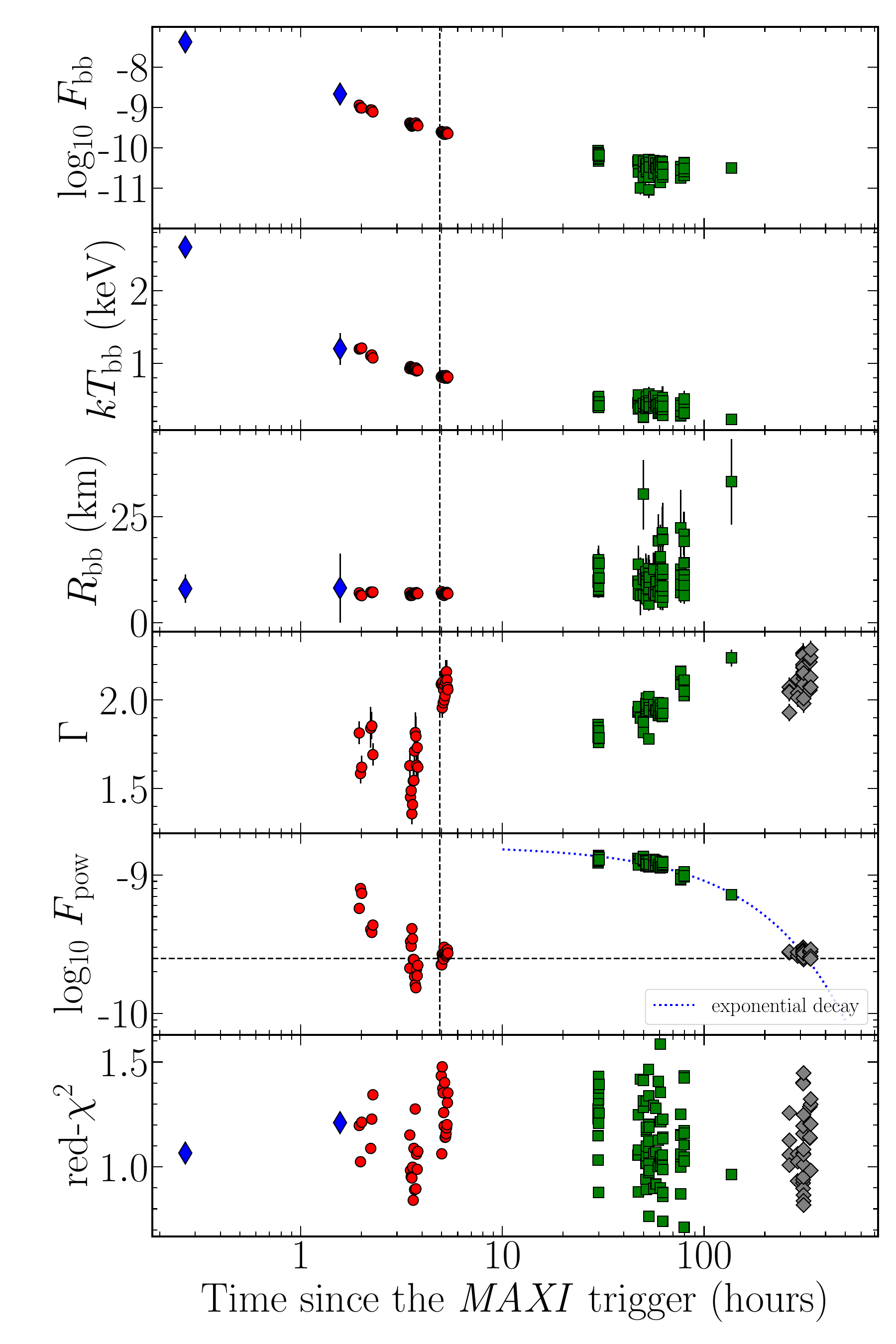}
\caption{The time-resolved spectroscopy of the long burst and outburst-like flare observed by MAXI and NICER. From top to bottom, we show the bolometric flux of \texttt{blackbody}, $F_\mathrm{ bb}$; the blackbody temperature, $kT_\mathrm{bb}$, the blackbody radius, $R_{\rm bb}$, which was calculated at a distance of 9.45 kpc; the power-law index, $\Gamma$; the bolometric flux of power-law, $F_\mathrm{pow}$; and the goodness of fit per degree of freedom, $\chi_{\nu}^{2}$. Blue diamonds represent MAXI survey data, while red circles denote NICER observations of the burst cooling tail (ObsID 3030080101). Green squares and gray diamonds represent the peak and decay phases of the outburst-like flare, respectively. The vertical dashed line marks the initial rise of the power-law flux and the early onset of the flare. 
}\label{fig:burst}
\end{figure}

\begin{figure}[ht!]
\includegraphics[width=\hsize]{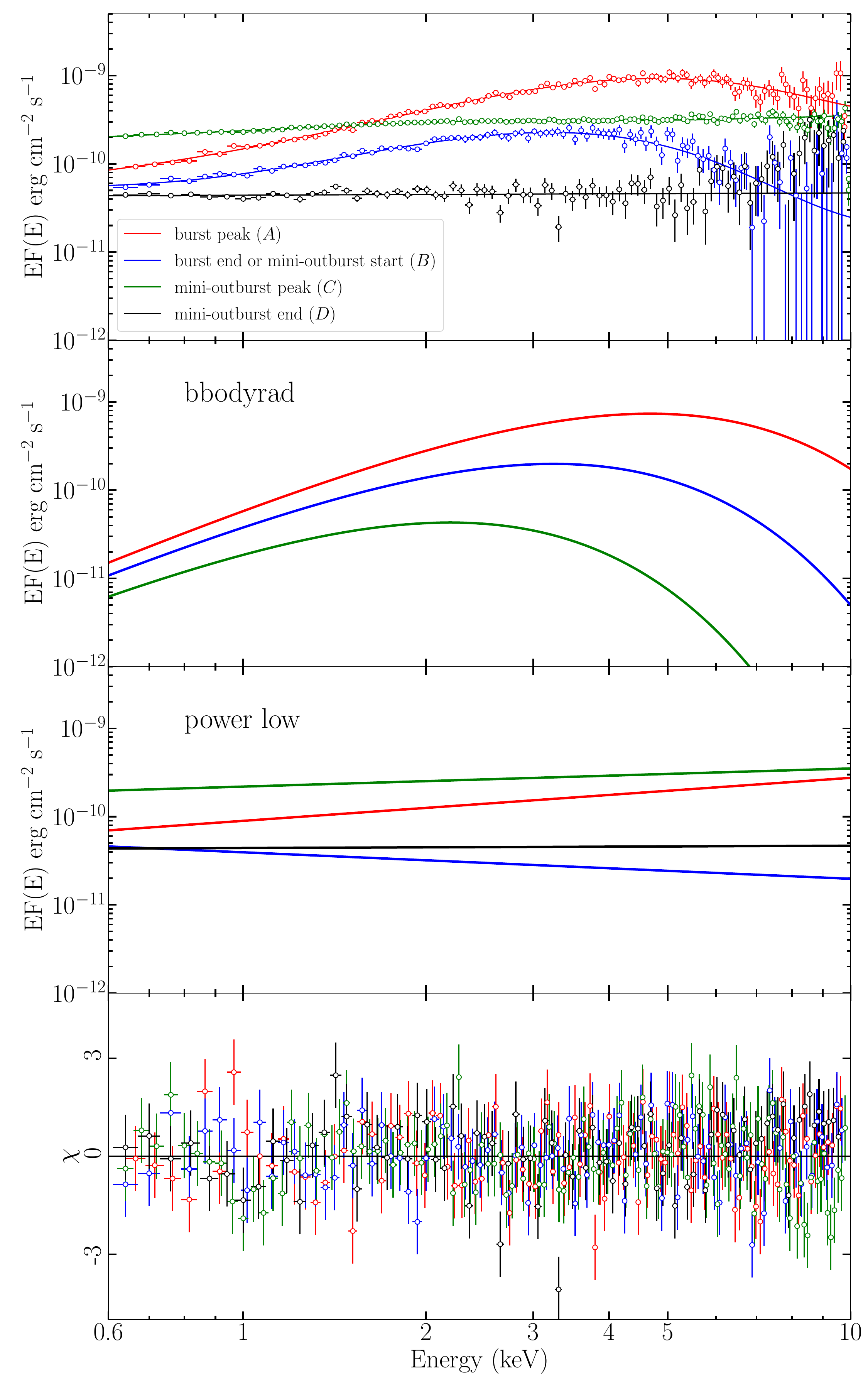}
    \caption{The absorbed best-fit spectra and the residuals in 0.5--10 keV obtained by fitting with the model \texttt{Tbabs*(bbodyrad+powerlaw)}. From top to bottom: each spectrum from regions A--D with the best-fit models (solid lines), the \texttt{blackbody} component, the \texttt{powerlaw} component, and residuals of the best-fit models to the spectra, respectively. The \texttt{blackbody} component are insignificant in the spectra from region D (outburst-like flare tail).
    }\label{fig:spec}
\end{figure}

This model fits the burst spectra well, yielding a reduced $\chi^2$ ($\chi^2_\nu$) $<1.5$. The best-fitting parameters and $\chi^2_\nu$ of the MAXI and NICER are shown in Fig.~\ref{fig:burst}.  We found $N_{\rm H} \approx 2.7 \times 10^{21}$~cm$^{-2}$, which is consistent across all spectra and is in excellent agreement with the results reported by \citet{Sanna2017A&A} and \citet{2020ATel}. The bolometric flux of the \texttt{bbodyrad} component decreased from an initial value of $1.1 \times 10^{-9}$~erg~cm$^{-2}$~s$^{-1}$ to $2.2 \times 10^{-10}$~erg~cm$^{-2}$~s$^{-1}$ during the observation. This corresponds to the blackbody temperature, $kT_{\rm bb}$, decreasing from $\approx1.2$ to $\approx0.8$~keV. Throughout this decay, the apparent blackbody radius, $R_{\rm bb}$, remained approximately constant at $\approx8$~km for the distance of 9.45 kpc. The initial 1.95~hr of the burst were not observed by NICER, indicating that the burst peak and any potential photospheric radius expansion phase might have been missed. Simultaneously, the persistent emission showed complex evolution. As the burst faded, the \texttt{power-law} flux initially decreased from $8 \times 10^{-10}$~erg~cm$^{-2}$~s$^{-1}$ to a minimum of $\sim(1-2) \times 10^{-10}$~erg~cm$^{-2}$~s$^{-1}$. During this time, the photon index $\Gamma$ exhibited nonmonotonic behavior, first decreasing from 1.8 to a harder value of 1.4, before softening again to 2.2. Notably, near the end of this observation, the power-law flux began to rise again. This rise lagged the changes in the photon index by approximately 1 hr, signaling the onset of the subsequent X-ray flare. The spectra and best-fitted model of intervals A--D, marked in Fig.~\ref{fig:lc_burst}, are shown in Fig.~\ref{fig:spec}.

\subsection{Spectral Analysis of the X-Ray Flare \label{subsec:outburst}}

During the X-ray flare, the spectra were initially modeled with an absorbed power law, \texttt{TBabs*powerlaw}. This model provided an adequate fit during the late decay phase of the flare, with $\chi^2_\nu \approx 1$. However, it fails near the peak of X-ray flare, yielding statistically unacceptable fits with $\chi^2_\nu$ up to 6. The addition of a blackbody component (\texttt{bbodyrad}) significantly improved the fit in the peak spectra; thus, the used model is \texttt{TBabs*(bbodyrad+powerlaw)}. The absorption column density was also fixed at $N_{\rm H} = 2.7 \times 10^{21}$~cm$^{-2}$, which was obtained in Sect.~\ref{subsec:burst}. The evolution of the best-fit parameters for the X-ray flare is shown as the green square and gray diamond in Fig~\ref{fig:burst}.

The X-ray flare was dominated by the power-law component. The photon index, $\Gamma$, softened from its preburst value of $\sim1.7$ to a peak value of $\approx2.2$, before rehardening to $\sim2.0$ during the decay. A weak thermal component was statistically required only near the outburst peak, with the bolometric flux $F_{\rm bb} < 9 \times 10^{-11}$~erg~s$^{-1}$~cm$^{-2}$, a blackbody temperature of $kT_{\rm bb} \sim 0.4$~keV, and an apparent radius of $\approx15$~km.

The evolution of the powerlaw and blackbody fluxes are shown in Fig~\ref{fig:burst}. The power-law flux rose by 1 order of magnitude from its preburst level of $\approx 2 \times 10^{-10}$~erg~s$^{-1}$~cm$^{-2}$ to a peak of $1.4 \times 10^{-9}$~erg~s$^{-1}$~cm$^{-2}$ approximately 29.8 hr after the burst trigger. The subsequent decay of the power-law flux is well described by an exponential decay. We estimated the total fluence of the X-ray flare to be $\approx 2.3 \times 10^{-7}$~erg~cm$^{-2}$, corresponding to a total energy release of $\approx 2.5 \times 10^{39}$~erg, assuming isotropic emission. The peak flux was well above the historical persistent flux in the range of $(2.0-5.0)\times 10^{-10}~{\rm erg~s^{-1}~cm^{-2}}$ \citep{0911_2019ATel_nicer, 0911_2021ATel_nicer}. To ensure the robustness of our spectral modeling, we systematically checked for statistical correlations among the free parameters (e.g., between the blackbody parameters and the power-law index or flux) across the flare observations. We found no strong or systematic parameter degeneracies.

\section{Discussion and conclusion} \label{sec:Dis}

In this work, we reported a long thermonuclear burst and subsequent outburst-like flare from MAXI J0911--655 observed by NICER and MAXI in 2020 August. We performed the spectral analysis of the burst and X-ray flare emissions. From the burst flux, we constrained the burst fuel in Sect.~\ref{sec:cooling}. We discuss the evolution of the power-law component to analyze the persistent emission during the burst and outburst-like flare in Sect.~\ref{sec:evo}. We propose that the disk instability caused by the long burst is responsible for the outburst-like X-ray flare, which is discussed in Sect.~\ref{sec:dim}.

\subsection{Burst parameters and burst fuel}
\label{sec:cooling}
\begin{figure}[ht!]
\includegraphics[width=\hsize]{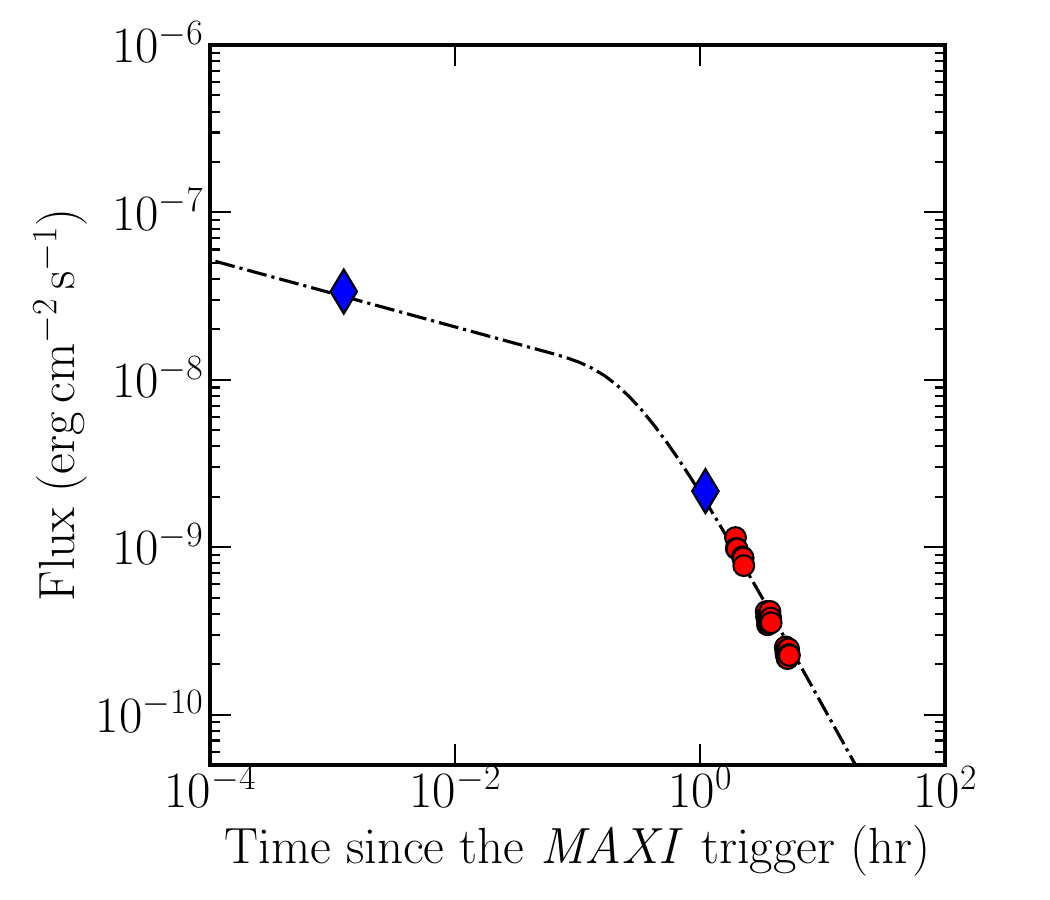}
\caption{The best-fitted burst decay flux by the model from \citet{Cumming04a}. We use the trigger time of MAXI as the burst peak time. The red points represent the data from
MAXI. 
} \label{fig:decay}
\end{figure}

As shown in Fig.~\ref{fig:decay}, the decay of the long X-ray burst bolometric flux can be well described by the analytic expression from \citet{Cumming04a} and \citet{Cumming06}, which depends on the energy release per unit mass, $E_{17}$, in units of 10$^{17}~{\rm erg~g^{-1}}$; and the ignition column depth, $y_{12}$, in units of 10$^{12}~{\rm g ~cm^{-2}}$. Because the true peak of the burst occurred before the NICER observation began, we assume the burst reached the empirical Eddington flux for a helium atmosphere, $F_\mathrm{Edd}\approx3.2\times10^{-8}~\mathrm{erg~cm^{-2}~s^{-1}}$ \citep{lewin1993x,Suleimanov17}. We estimated  $E_{17}\sim1.2$ and $y_{12}\sim0.1$, and found the burst peak is close to the trigger time from MAXI. 

The burst fluence can be calculated from the estimated ignition column depth $y_{\rm ign}$,
\begin{align}
    f_{\rm b}=\frac{4\pi y_{\rm ign}R^{2}_{\rm NS}Q_{\rm nuc}}{4\pi d^{2}(1+z)},
\end{align}
where $R_{\rm NS}= 10$~km, $Q_{\rm nuc} \approx 1.31 ~{\rm MeV~nucleon^{-1}}$, the source distance $d = 9.45$~kpc, and the gravitational redshift on the NS surface $z=0.31$ for the NS mass of $M_{\rm NS}=1.4 M_{\odot}$.
We obtained the burst fluence $f_{\rm b}\approx 1.1\times 10^{-4}~{\rm erg~cm^{-2}}$, and the total energy release of $\approx1.2\times10^{42}$ erg. From these values, we can calculate a characteristic decay time $\tau=f_{\rm b}/F_{\rm peak} \approx 0.92$ hr, which is longer than the $\tau = 0.72$~hr timescale derived independently from the exponential fit to the NICER light curve.

The local accretion rate can be calculated from its preburst emission \citep{Galloway08} as the following equation: 
\begin{equation}
\begin{split}
    \Dot{m}
    &=\frac{L_\mathrm{ {per}}(1+z)}{ 4\pi R_{\rm NS}^{2}(GM_{\rm NS}/R_{\rm NS})}\\
    &\approx 1.7\times 10^{3}\biggl(\frac{F_\mathrm{{per}}}{4\times10^{-10}\mathrm{~ergs ~ cm^{-2}~s^{-1}}}\biggr)\biggl(\frac{d}{8\mathrm{~ kpc}}\biggr)^{2}\\
    &\quad\times\ \biggl(\frac{M_{\rm NS}}{1.4M_{\odot}}\biggr)^{-1}\biggl(\frac{1+z}{1.31} \biggr)\biggl(\frac{R\mathrm{_{NS}}}{10\mathrm{~ km}}\biggr)^{-1}\mathrm{g~cm^{-2}}\mathrm{~s^{-1}},\label{eq:lo_accration}
\end{split}      
\end{equation}\\ 
where $F_{\rm per}$ is the persistent flux. Lacking a preburst NICER observation, we assume the persistent flux is equal to the power-law flux measured at the end of the burst tail, $F_{\rm per} \approx 2.8 \times 10^{-10} {\rm ~erg~cm^{-2}~s^{-1}}$. This flux corresponds to a local accretion rate of $\dot{m} = (1.7 \pm 0.01) \times 10^3~{\rm g~cm^{-2}~s^{-1}}$ for a distance of $9.45$ kpc \citep{01Paltrinieri}. This places the source at an accretion rate of $\dot{m} \sim 0.01\dot{m}_{\rm Edd}$, where we assume the local Eddington accretion rate is $\dot{m}_{\rm Edd}=(8.8\times10^4)[1.7/(X+1)]~{\rm ~g~cm^{-2}~s^{-1}}$, and the accreted hydrogen fraction $X=0$ for the UCXB source MAXI~J0911--655.

Using the derived ignition depth and accretion rate, we can predict the burst recurrence time via $\Delta t_{\mathrm{rec}}=(y_{\mathrm{ign}} / \dot{m})(1+z) \approx 2.45$~yr. This prediction is remarkably close to the actual observed recurrence time of $\Delta t_{\rm rec} \approx 2.17$~yr to the next long burst detected on 2022 July 22. This strong agreement further validates our derived burst parameters.

Similar to the long burst from 4U 1850--087 reported by \citet{Lu24}, the burst duration $\sim0.92$ hr falls between the longest intermediate-duration bursts and the shortest superbursts. The combination of a long duration ($\sim0.8$~hr), a low accretion rate ($\sim0.01\dot{m}_{\rm Edd}$), and the ultracompact nature of the system (implying a helium-rich donor; \citealt{Sanna2017A&A}) strongly suggests this was an intermediate-duration burst fueled by unstable helium burning. To test this hypothesis, we calculate the $\alpha$-parameter, the ratio of integrated persistent energy to the burst energy. We measure an observed value of $\alpha = (F_{\rm pers} \times \Delta T_{\rm rec}) / f_{\rm b} \sim 174$. This is in good agreement with the theoretical value of $\alpha=44~M_\mathrm{NS}R_\mathrm{NS}^{-1}(Q_\mathrm{nuc}/4.4\mathrm{MeV/nucleon})^{-1}\approx121$  for pure helium burning \citep{Falanga08}. The consistency between our observed value and theory provides compelling evidence that this event was indeed a powerful, helium-fueled, intermediate-duration burst.

\subsection{Evolution of the power-law emission during the long burst}
\label{sec:evo}
During the cooling of the long burst, we adopted a power-law component to describe the persistent emission.
Initially, the NICER spectra  showed a low power-law photon index $\sim 1.5$, and a rapidly decaying flux. The power-law flux decreased from $7.98\times 10^{-10}$ to $1.45\times 10^{-10}$ $~{\rm erg~s^{-1}~cm^{-2}}$, lower than the average level, i.e., $1.53\times 10^{-10}~{\rm erg~s^{-1}~cm^{-2}}$, within 3.8 hr.  During the low-flux stage, the photon index of the power law drops to minimum $\sim 1.4$.  Finally, after exiting the low-flux stage, the power law becomes more prominent, the photon index softens to 2.2, and the flux recovers its persistent level. See Fig.~\ref{fig:burst}.

The impact of the burst emission on the mass accretion flow could account for the power-law component evolution. The PR effect could drag the inner disk material onto the NS surface \citep{Ballantyne2005ApJ,Fragile20}, causing the enhancement of the persistent emission. Subsequently, the burst radiation could prevent the inward migration of the accretion disk material \citep{Ballantyne2005ApJ,int2011A&A}.  Following  the bright phase of the burst, one might expect that it takes some time for the material to refill the inner accretion disk \citep{Bult21,Peng_2025}. As a consequence, the persistent emission was lower than the preburst level. After a superburst from 4U 1820--30, \citet{Peng_2025} estimated the viscous timescale of the recovery of the accretion disk from $1000r_g$ around 1.8 hr. 

Another aspect is that the early X-ray burst has a lower power-law index than the later stage. This result indicates that the geometric configuration of the Comptonization region differs fundamentally between the bright burst phase and the quiescent persistent emission state. To interpret the complex evolution of the power-law component, we consider the dynamic response of the corona to the intense burst radiation. As discussed by \citet{Bult21} in the context of IGR J17062--6143, an energetic burst floods the local environment with soft X-ray photons, which can rapidly cool the hot coronal plasma through inverse Compton scattering \citep[see also][]{Speicher20}.Furthermore, the burst-induced enhancement of the disk mass accretion rate would provide a temporary increase in coronal heating efficiency, which in turn partially counteracts the strong radiative cooling of the corona during the burst decay phase. This intense radiative cooling is expected to drive the corona to physically condense. Such a structural collapse naturally explains the sharp decrease in the power-law flux and the corresponding spectral hardening (i.e., the drop in the photon index to $\Gamma \sim 1.4$) that we observe during the early decay phase of the burst. As the burst emission decays, this intense radiative pressure is lifted. During this recovery phase, the expanding, low-density corona may still be subjected to residual cooling from the fading burst. This combination of lower optical depth and suppressed electron temperature reduces the efficiency of inverse Compton scattering, accounting for the transient spectral softening ($\Gamma \sim 2.2$) observed as the burst cooling tail fades.

\subsection{Long Thermonuclear Burst Trigger an Outburst-like  X-Ray Flare}
\label{sec:dim}

From our observation, we found a rebrightening of the persistent emission starting about 30~hr after the onset of the long burst, which developed into a clear outburst-like X-ray flare (Fig.~\ref{fig:lc_burst}). The flare showed a FRED profile, characteristic of LMXB outbursts \citep{Wood_2001,Lipunova_2015}. The historical light curve (Fig.~\ref{fig:lc}) shows that for more than 1 yr prior to the 2020 long burst, the source remained in a low/hard state and there was no significant increase in the accretion rate. Therefore, the possibility of this outburst being triggered by an external change in the accretion rate can be safely ruled out. We propose that the flare is a direct consequence of the long burst: intense irradiation from the burst heats the inner accretion disk, enhances the ongoing thermal-viscous accretion processes, and launches an inside-out heating front that leads to a temporary increase in the mass accretion rate.

\begin{figure*}[ht!]
\includegraphics[width=\hsize]{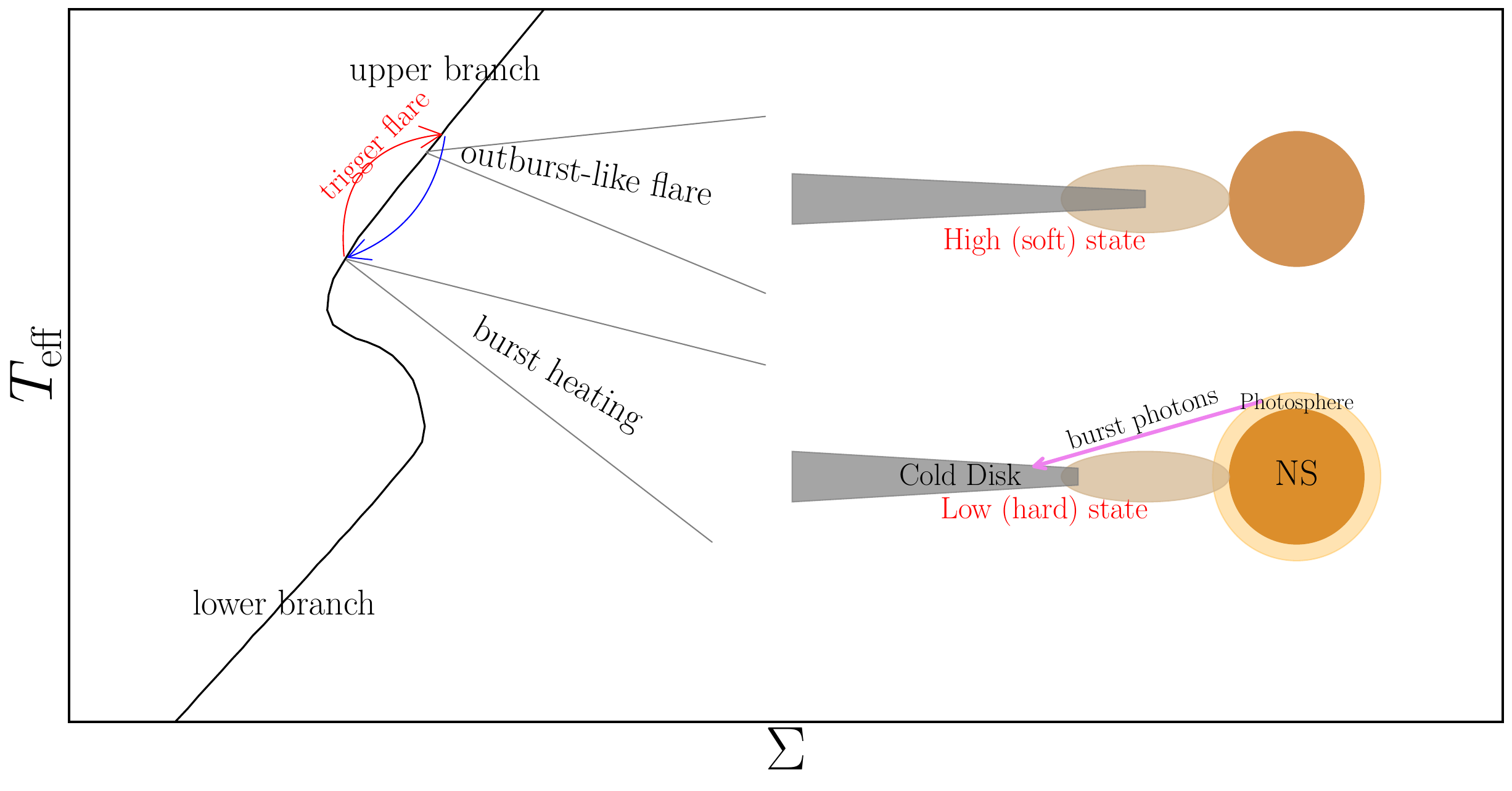}
\caption{Schematic diagram of the burst-triggered disk instability. \textit{Left:} thermal equilibrium S-curves for the accretion disk, illustrating effective temperature ($T_{\rm eff}$) versus local disk mass. The solid, dashed, and dashed-dotted lines represent increasing values of the irradiation parameter, $C$. Prior to the flare, the disk is already in an active, hot state on the upper branch, corresponding to the source's persistent low-luminosity outburst. Intense X-ray heating from the thermonuclear burst effectively increases $C$ and drives the disk to an even hotter, higher-accretion state further up the upper branch, as indicated by the red arrow. As the flare decays, the disk cools and returns to its previous persistent low-luminosity state, indicated by the blue arrow. \textit{Right:} the corresponding physical geometry of the system. In the preflare active state (bottom, ``Low (hard) state''), photons from the thermonuclear burst strongly irradiate the accretion disk. This heating triggers an inside-out heating front that transitions the disk into an even hotter, more highly viscous state (top, ``High (soft) state''). Consequently, the inner disk radius moves inward, enhancing the accretion rate and producing the observed outburst-like X-ray flare.} \label{fig:Schematic_picture}
\end{figure*}

The theoretical feasibility of this mechanism depends on the energy deposited by the burst. MAXI J0911--655 accreted at a very low persistent rate of $\dot{m}\approx0.01\dot{m}_{\mathrm{Edd}}$ (Sect.~\ref{sec:cooling}). The long burst had a duration $\tau>0.7$~hr (43~minutes, Sect.~\ref{subsec:burst_lc}), which is 2 orders of magnitude longer than the thermal timescale of the inner disk, $t_{\mathrm{th}}\sim4$ minutes \citep[see, e.g.,][]{Ballantyne2005ApJ}. Consequently, the burst radiation can deposit a significant amount of energy into the disk. The irradiation temperature of the disk can be expressed as
\begin{equation}
\begin{split}
\sigma T_{\mathrm{irr}}^{4}=C\frac{L_{X}}{4\pi R^{2}},
\end{split}      
\end{equation}
where $L_{X}$ is the X-ray luminosity of the central source,  $C$  represents the fraction of the X-ray luminosity, and $R$ is the radius of the disk. While $C\approx0.005$ for a persistent low/hard state \citep{Dubus2001A&A}, the luminosity approaches the Eddington limit during the burst, increasing the effective $C$ by a factor of $\sim$ 100 \citep{Kuulkers2009A&A}.  Numerical simulations by \citet{Fragile20} show that such a strong irradiation raises the density-weighted temperature of the disk by about 0.5 orders of magnitude over a wide range of radii.

In the framework of the DIM, strong external irradiation significantly modifies the disk's thermal equilibrium and accretion dynamics \citep{Dubus2001A&A,Lasota2008}. The sequence of events under this framework is schematically illustrated in Fig.~\ref{fig:Schematic_picture}. Prior to the long burst, the source is in a persistent, low-luminosity active state; therefore, the accretion disk already resides on the upper (hot, ionized) branch of the S-curve, corresponding to a low/hard state with a truncated inner disk. Because the disk is already in this active state, the burst irradiation acts as an intense, transient external heat source that amplifies the existing hot state. It pushes the already hot inner disk to an even hotter, higher‑accretion regime. As the inner disk becomes hotter, the turbulent viscosity ($\alpha$) increases dramatically, driving a secondary heating front that propagates outward. This inside-out front is opposite to the usual outside-in outbursts seen in transient LMXBs, but it is exactly what is expected when the central source experiences a sudden brightening \citep{Serino2012,Fragile20,Speicher2024ApJ}.

To quantitatively match the postburst accretion rate, the turbulent viscosity parameter must increase. In the preburst steady state, the observed accretion rate is well reproduced with $\alpha = 0.1$ using the standard viscosity prescription $\nu = \alpha c_s H$, $c_{s}$ is the sound speed, and $H$ the local disk scale height. Following the burst, as the disk temperature rises and the scale height nearly doubles \citep{Fragile20}, maintaining $\alpha = 0.1$ causes the predicted mass accretion rate $\dot{M} = 3\pi \Sigma \nu$ to fall significantly below the observed peak value. Matching the peak accretion rate therefore requires $\alpha \approx 0.2$. This factor-of-two enhancement in the viscosity parameter is qualitatively consistent with the temperature-dependent activation of magnetorotational turbulence \citep{Fragile20}.

Using this enhanced viscosity, the time required for the heating front to travel from the innermost radius ($R_{\mathrm{in}}\sim10^{9}$~cm) to the outer edge of the heated region ($R_{\mathrm{out}}\sim5\times10^{9}$~cm) is governed by the viscous time scale \citep{Dubus2001A&A}:
\begin{equation}
\begin{split}
t_{\mathrm{vis}}=\frac{R^{2}}{\nu}= (GM_{\mathrm{NS}}R)^{1/2}\frac{\mu m_{\mathrm{H}}}{\alpha_{\mathrm{vis}}kT_{\mathrm{c}}},
\end{split}      
\end{equation}\\
where $T_{\mathrm{c}}$ is the midplane temperature and $\mu=1.33$ (fully ionized pure helium) and $\alpha_{\mathrm{vis}}=0.2$ are typical values. With $T_{\mathrm{c}}\approx7.4\times10^{5}$~K (appropriate for a heated disk), we obtain $t_{\mathrm{vis}}\approx29$~hr. This theoretical timescale is in excellent agreement with the observed 30 hr rise time of the outburst-like flare, strongly supporting the interpretation that the flare is caused by the delayed arrival of disk material set into motion by the burst-amplified heating front.

\begin{figure}[ht!]
\includegraphics[width=\hsize]{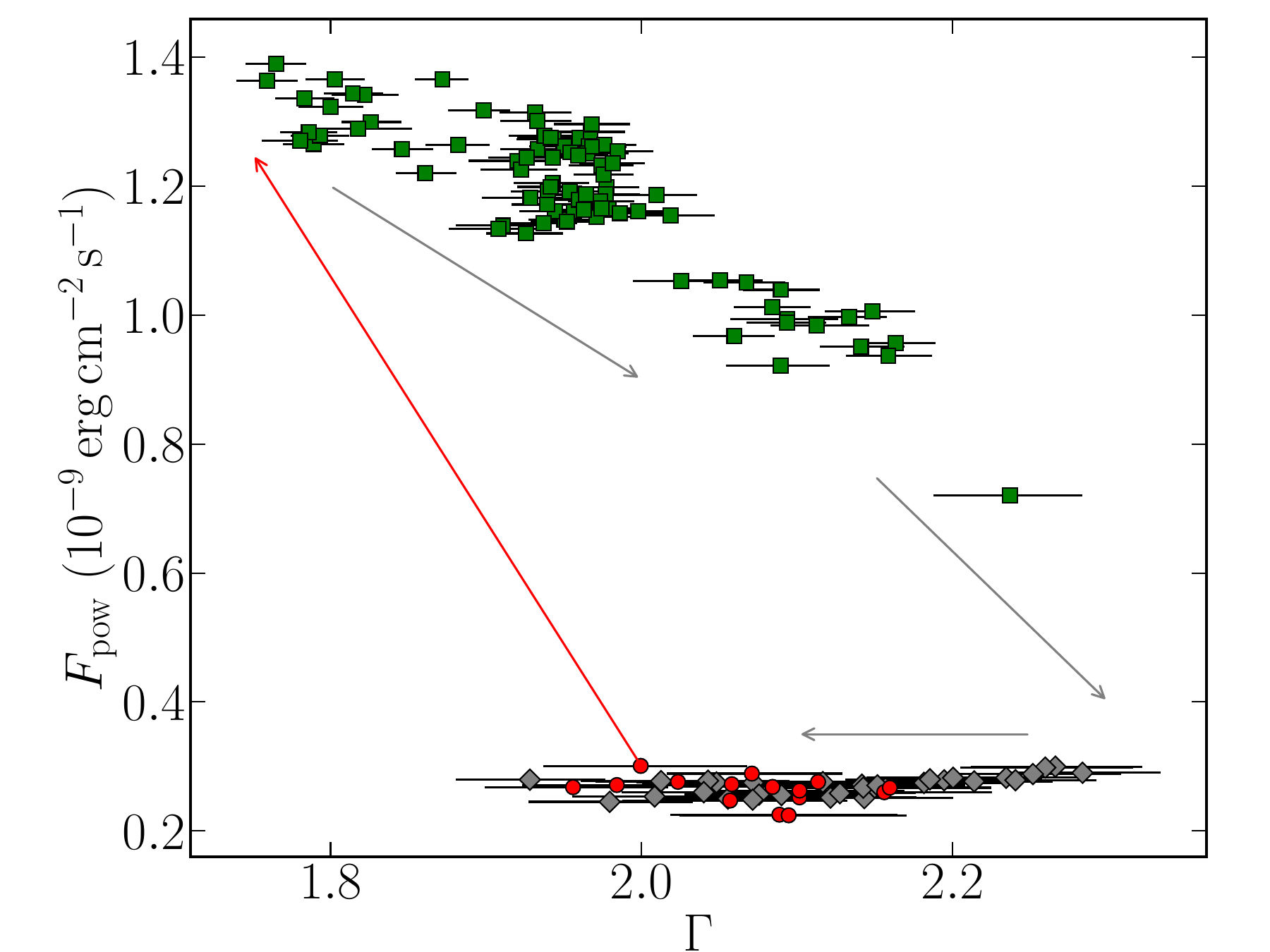}
\caption{Relations of power-law flux vs. the index $\Gamma$. The arrow indicates the direction of evolution.
Red arrow: burst rise phase---the source evolves from a low/hard state (low flux, hard spectrum) to a high/soft state (high flux, soft spectrum).  
Gray arrow: outburst decay phase----the source returns from the high/soft state to the low/hard state (flux declines, spectrum hardens). 
The complete cycle illustrates the typical hysteresis behavior of a NS LMXB outburst.} \label{fig:flux_vs_ga}
\end{figure}

The spectral evolution of the source provides further observational evidence for this accretion-driven state transition (Fig.~\ref{fig:flux_vs_ga}). During the flare, the power-law photon index $\Gamma$ softens from $\sim1.7$ to $\sim2.2$ (Fig.~\ref{fig:flux_vs_ga}), indicating that the corona is cooled by the increased flux of soft photons emitted from the encroaching inner disk \citep{Maccarone03}. At the peak of the flare, a weak blackbody component emerges with $kT_{\mathrm{bb}}\approx0.4$~keV and an apparent radius of $\approx15$~km. This radius is larger than the NS surface, confirming it corresponds to the inner region of the accretion disk, which has moved inward and become hot enough to emit thermal radiation. 

While Fig.~\ref{fig:flux_vs_ga} primarily traces the descending (postpeak) phase of the outburst due to observational gaps during the initial rise, the available data reveal an upward trend in the power-law flux beginning 5 hr after the onset of the thermonuclear burst, while the spectrum still resided in the low/hard state. The source subsequently evolves from a high-$F_{\rm pow}$, low-$\Gamma$ state near the soft-state peak to a low-$F_{\rm pow}$, high-$\Gamma$ state during the soft-to-hard transition. After the peak, the blackbody fades, the photon index returns to $\sim2.0$, and the disk cools and recedes, returning the source to its preburst low/hard state on the upper branch of the S-curve. This complete cycle and the suppression of the power-law component during the initial hardening displays the characteristic hysteresis behavior expected of standard NS LMXB outbursts.

The burst-driven X-ray flare in MAXI J0911--655 stands in stark contrast to the post-burst behavior observed in IGR J17062--6143 \citep{Bult21}. Like MAXI J0911--655, IGR J17062--6143 is an AMXP in an ultracompact binary that experienced an intermediate-duration burst. However, following its burst, IGR J17062--6143 exhibited a prolonged three-day intensity dip, which \citet{Bult21} interpreted as a temporary suppression of the accretion rate due to the evacuation and gradual viscous refilling of the inner disk. Similarly, \citet{Peng_2025} reported the quenching of the persistent emission in 4U 1820--30 during a superburst. These differences demonstrate that burst irradiation can drive competing physical processes. While radiation pressure or PR drag can evacuate the inner disk under certain conditions \citep{Fragile20}, the extreme heating can alternatively amplify the local viscosity and trigger a thermal-viscous heating front, as observed in MAXI J0911--655. The parameters determining which of these two pathways (disk depletion versus instability enhancement) dominates likely intricately depend on the specific burst energetics and the preburst disk density profile.

\section{Summary}\label{sec:sum}

In this work, we detected a long-duration burst and succeeding outburst-like flare from MAXI J0911--655 using joint NICER and MAXI observations. These observations provide a unique opportunity to investigate the impact of a powerful thermonuclear burst on the instability of the accretion disk. From the time-resolved burst spectroscopy during the burst decay, we obtained the burst decay time, $\tau \approx 43$ minutes, the burst fluence, $E_\mathrm{b} \approx 8.1 \times 10^{41}$ erg. The observations show that the source has been observed still in the low/hard state with a cold disk truncated on the outside, which could be heated by the burst and increased the temperature by about 0.5 order of magnitude \citep{Fragile20,Speicher2024ApJ}. As the trigger of the outburst-like flare, we observed that the spectrum of the source had softened and the inner accretion was gradually restored to its previous state (low/hard state). Therefore, we propose a possibility that the instability of the accretion disk caused by the long-duration burst is responsible for the outburst-like X-ray flare.

\begin{acknowledgments}
We appreciate the referee for the valuable comments and suggestions, which improved the manuscript. This work was supported by the Major Science and Technology Program of Xinjiang Uygur Autonomous Region (No. 2022A03013-3). Z.S.L. and Y.Y.P. were supported by National Natural Science Foundation of China (12273030, 12103042). This research
was supported by the International Space Science Institute
(ISSI) in Bern and the International Space Science Institute–
Beijing (ISSI-BJ), through the joint ISSI/ISSI-BJ International
Team project ``Thermonuclear X-ray Bursts: From Simulations
to Multi-Wavelength Observations" led by Z. Li and
D. Galloway (ISSI Team project \#25-647). This work made use of data from the High Energy Astrophysics Science Archive Research Center (HEASARC), provided by NASA’s Goddard Space Flight Center. 

\end{acknowledgments}

\vspace{5mm}

\bibliography{name}{}
\bibliographystyle{aasjournal}

\end{document}